\documentclass[11pt]{article}

\usepackage{jcappub}

\usepackage{amssymb,amsfonts,amsmath,graphicx,}
\usepackage{color}
\usepackage{enumerate}
\usepackage{braket}
\usepackage{epstopdf}
\usepackage{bm}
\usepackage{latexsym}
\usepackage[normalem]{ulem}
\usepackage{graphicx}
\usepackage{color}
\usepackage{subcaption}

\makeatletter
\gdef\@fpheader{}
\makeatother

\newcommand{\dis}[1]{\begin{equation}\begin{split}#1\end{split}\end{equation}}
\newcommand{\etal}{et al.\,}

\newcommand{\be}{\begin{equation}}
\newcommand{\ee}{\end{equation}}
\newcommand{\bea}{\begin{eqnarray}}
\newcommand{\eea}{\end{eqnarray}}

\def\PR{\mathcal{P_{\mathcal{R}}}}
\def\ie{{\it i.e.}}

\def\nn{\nonumber} 

\def\f{\frac}
\def\l{\left}
\def\r{\right}
\def\d{{\rm d}}
\def\s{{\sigma}}

\newcommand{\VEV}[1]{\langle #1 \rangle}
\newcommand{\bfrac}[2]{{\left(\frac{#1}{#2} \right)  }}
\newcommand{\eq}[1]{Eq.~(\ref{#1})}

\newcommand\mpc{\,{\rm Mpc}}

\newcommand{\Mp}{M_{\rm P}}

\newcommand{\cR}{\mathcal{R}}

\begin{document}
\title{Reconstruction of potentials of the hybrid inflation in the light of primordial black hole formation}
\author[a]{Ki-Young Choi,}
\emailAdd{kiyoungchoi@skku.edu}
\affiliation[a]{Department of Physics, Sungkyunkwan University,  16419 Korea}
\author[b]{Su-beom Kang,}
\affiliation[b]{Sungkyunkwan University,  16419 Korea}
\emailAdd{subeom527@gmail.com}
\author[a]{and Rathul Nath Raveendran}
\emailAdd{rathulnath.r@gmail.com}

\date{today} 
\abstract{
The large enhancement of the primordial power spectrum of the curvature perturbation can seed the formation of primordial black hole,
that can play as a dark matter component in the Universe.
In multi-filed inflation models, the curved trajectory of the scalar fields in the field space can generate a peak in the power spectrum on small scales due to the existence of the isocurvature perturbation. Here we show that a potential can be reconstructed from a given power spectrum, which is made of a scale-invariant one on large scales and the other function with a peak on small scales. In multi-field inflation models the reconstructed potential may not be unique and we can find different potentials from a given power spectrum.
}
\maketitle

\section{Introduction}
\label{intro}
The observation of the large scale structures and the anisotropy of the cosmic microwave background (CMB) allowed the precise determination of the primordial power spectrum at large scales~\cite{Akrami:2018odb}. Cosmic inflation can naturally produce the required spectrum in addition to resolving the problems in the standard big bang model. However on small scales, the power spectrum 
is constrained weakly only by the formation of primordial black hole (PBH), ultracompact minihalo, or dark matter (DM) annihilation for some specific models~\cite{Bringmann:2011ut,Choi:2015yma}. 

The PBHs can form through the collapse of large fluctuations of the density in the early Universe when a given scale enters the horizon~\cite{Zeldovich:1967lct,Hawking:1971ei,Carr:1975qj}. The PBHs can evaporate via Hawking radiation, however those with the mass larger than  $5\times 10^{14}$g can survive until today. The present remnant of PBHs can contribute to the  non-baryonic component of dark matter~\cite{Page:1976df}.  See the recent review~\cite{Green:2020jor} for PBH as a candidates for dark matter.

For the formation of PBH, typically the amplitude of the power spectrum need to be larger than around $10^{-2}$, that is $10^7$ bigger than that at the CMB scales, ${\mathcal P}_{\rm CMB}\sim  10^{-9}$.
Several models were suggested to generate the large enhancement  at small scales with a single scalar field, such as features in the potential, running mass of the inflaton, hilltop models, and inflection point~\cite{Ivanov:1994pa,Stewart:1996ey,Kohri:2007gq,Kohri:2007qn,Alabidi:2009bk,Garcia-Bellido:2017mdw,Germani:2017bcs,Motohashi:2017kbs,Ballesteros:2017fsr}. However, for a single field inflation, the slow-roll must be violated at scales between the CMB scale and PBH mass scales. In this case, a numerical calculation is needed  to evaluate the power spectrum properly.

In Ref.~\cite{Hertzberg:2017dkh}, Hertzberg \etal provided a method to reconstruct the  inflaton potential  from a given power spectrum within canonical singled field model. They applied this method to the formation of PBHs and confirmed again that the slow-roll conditions need to be violated in order to generate a significant spike in the spectrum.

The multi-field models which can generate PBHs include hybrid inflation, double inflation, and a curvaton field~\cite{Yokoyama:1995ex,Kawasaki:1997ju,Yokoyama:1998pt,Kawasaki:2012wr,Clesse:2015wea,Fumagalli:2020adf,Braglia:2020eai}.
For non-canonical kinetic terms, the accurate analytic prediction was derived for the formation of PBHs in multifiled case~\cite{Palma:2020ejf}.

In this paper, we propose a method to reconstruct a potential of multi-field scalar model from the given power spectrum. We consider a primordial power spectrum made of a scale-invariant one on large scales and the other with a peak on small scales in the light of generating PBHs. We use a hybrid-type potential and show that a few different examples of the reconstructed potential. We find that the reconstructed potential may not be unique.
Once we know the potential, it is possible to obtain the power spectrum numerically by solving the exact equations of motions. Lastly, we present these numerical results to support our analytical arguments.

In Sec.~\ref{back}, we summarize the background evolution and the perturbations during inflation with  two scalar fields, and in Sec.~\ref{deltaN} we give the formulation of the $\delta N$ formalism approximation. In Sec.~\ref{recon}, we introduce a method to reconstruct the potential from a given power spectrum and show the analytical results with numerical calculation.
We conclude in Sec.~\ref{con}.

\section{Background and perturbations of two-field  model}
\label{back}
In this section, we briefly revisit the evolutions of relevant background quantities and scalar perturbations (for more details, see~ \cite{Lalak:2007vi,Gordon:2000hv}).
We shall focus on models with two real scalar fields, $\phi$ and $\psi$, 
described by the action
\begin{equation}
S[\phi_I] 
= -\int \d^4 x\, \sqrt{- g }\, 
\l[\frac{1}{2}\,\sum_{I=1}^2 \partial_{\mu}\phi_I\,\partial^{\mu}\phi_I 
+ V(\phi_I)\r],
\end{equation}
where $\phi_I=\{\phi,\psi\}$.
The equations of motion of the fields  can be written as
\begin{equation}
{\ddot \phi_I}+3\, H\, {\dot \phi}_I + V_{\phi_I}=0,
\end{equation}
where $H={\dot a}/a$ is the Hubble parameter and $V_{\phi_I}=\d V/\d\phi_I$.
Two Friedmann equations describing the evolution of the scale factor are 
given by
\begin{subequations}
\begin{eqnarray}
H^2 &=& \frac{1}{3\,\Mp^2}\, 
\l[\frac{1}{2}\, \l({\dot \phi}^2 + {\dot \psi}^2\r)+ V(\phi,\psi)\r],\\
{\dot H} &=& -\frac{1}{2\,\Mp^2}\,\l({\dot \phi}^2 + {\dot \psi}^2\r).
\end{eqnarray}
\end{subequations}
In terms of the e-folding number, $N_t$, defined as $N_t={\rm ln}\,(a/a_i)$, 
where $a_i$ is the scale factor at a suitably chosen time, the field equations are
\dis{
\frac{d^2 \phi}{dN_t^2} + \l(3- \epsilon_{H}\r)\, \frac{d \phi}{dN_t} + \frac{V_{\phi}}{H^2}=0,\\
\frac{d^2 \psi}{dN_t^2} + \l(3-\epsilon_{H} \r)\, \frac{d \psi}{dN_t} + \frac{V_{\psi}}{H^2}=0,
\label{eomN}
}
where $\epsilon_{H}\equiv -\f{1}{H} \frac{dH}{dN_t}$ is the Hubble slow roll parameter.

Since there are two fields involved, evidently, apart from the curvature 
perturbation, isocurvature perturbation also arises. 
In the spatially flat gauge, for instance, the Mukhanov-Sasaki 
variables associated with the curvature and the 
isocurvature perturbations $v_\sigma$ and $v_s$ are given by
\begin{eqnarray}
v_\sigma&=&a\,\l({\rm cos}\,\theta\,\delta\phi
+{\rm sin}\,\theta\,\delta\psi\r),\\
v_s&=&a\,\l(-{\rm sin}\,\theta\,\delta\phi
+{\rm cos}\,\theta\,\delta\psi\r),
\end{eqnarray}
where ${\rm \cos}\,\theta=\dot{\phi}/\dot{\sigma}$,
${\rm sin}\,\theta=\, \dot{\psi}/\dot{\sigma}$ and
$\dot{\sigma}^2= \dot{\phi}^2+\dot{\psi}^2$.
The curvature and the isocurvature perturbations are defined
as ${\cal R}=v_\sigma/z$ and ${\cal S}=v_s/z$, respectively,
with $z=a\,\dot{\sigma}/H$~\cite{Lalak:2007vi}.

\par

It is convenient to introduce the adiabatic and entropy vectors
$E_\sigma^I$ and $E_s^I$ in the field space, defined 
as 
\begin{eqnarray}
E_\sigma^I&=&({\rm cos}\,\theta,
{\rm sin}\,\theta),\\
E_s^I&=&(-{\rm sin}\,\theta,
{\rm cos}\,\theta),
\end{eqnarray}
 where $I=\{\phi,\psi\}$.
The equations governing the gauge invariant 
Mukhanov-Sasaki variables $v_\sigma$ and
$v_s$ can be expressed as~\cite{Lalak:2007vi}
\begin{subequations}
\label{eq:ms}
	\begin{eqnarray} 
		v_{\sigma}'' 
		+\l(k ^2 - \f{z''}{z}\r)\, v_{\sigma}
		&= &\frac{1}{z}\, \l(z\, \xi\, v_s\r)',\\
		v_s'' + \l(k^2- \f{a''}{a}+a^2\, \mu_{s}^2\r)\, v_{s}
		&=&-z\, \xi\, \l(\f{v_\sigma }{z}\r)',
	\end{eqnarray}
\end{subequations}
where $\xi= -2\,a\,V_s/{\dot{\sigma}}$ and the quantity $\mu_s^2$
is given by
\begin{eqnarray}
	\mu _{s} ^2 
	& = & V_{ss} -\l(\frac{V_s}{\dot{\sigma}}\r)^2,
\end{eqnarray}
with the subscript $\phi$ or $\psi$ indicating differentiation
with respect to the fields.
Also, the quantities $V_\sigma$, $V_s$ and $V_{ss}$ are given by 
$V_\sigma=E_\sigma^I\,V_I$, $V_s=E_s^I\,V_I$ and $V_{ss}=E_s^I\, 
E_s^J\,V_{IJ}$, with implicit summations assumed over the repeated 
indices~$I$ and $J$. 

As we know, the perturbations considered are quantum in nature. We can quantise the perturbations by promoting the variables to quantum operators as~\cite{Lalak:2007vi}
\bea
\hat{v}_{\s}&=& f_{\s} \hat{a} + f_{\s}^\star \hat{a}^{\dagger}+g_{\s} \hat{b} + g_{\s}^\star \hat{b}^{\dagger},  \\
\hat{v}_{s}&=& f_{s} \hat{a} + f_{s}^\star \hat{a}^{\dagger}+g_{s} \hat{b} + g_{s}^\star \hat{b}^{\dagger},
\eea
where $f_{\s, s}$ and $g_{\s, s}$ are the solutions of \eq{eq:ms} , $(\hat{a}, \hat{b}) $ and $(\hat{a}^\dagger, \hat{b}^\dagger) $ are the annihilation and creation operators. Vacuum states are defined as
\be
\hat{a} \rvert 0 \rangle = \hat{b} \rvert 0 \rangle = 0.
\ee
When the modes are very deep inside the Hubble radius, the equations of motion governing the set of variable ($f_\s ,f_s$) and ($g_\s, g_s$) are decoupled and we shall set the initial conditions, as usual, by the Minkowski-like vacuum  as
\begin{eqnarray}
	f_{\s}(\eta)&=& g_{s}(\eta)=\frac{e^{-i k \eta}}{\sqrt{2 k}}~,\\
	f_{s}(\eta)&=& g_{\s}(\eta)=0~.
\end{eqnarray}
The two scalar power spectra can be expressed as~\cite{Lalak:2007vi,GarciaBellido:1995qq}
\dis{
	\PR &=  \f{k^3}{2 \pi^2}  \f{\vert f_{\sigma}\vert^2 + \vert g_{\sigma}\vert^2 }{z^2}, \\
P_S &=  \f{k^3}{2 \pi^2}  \f{\vert f_s\vert^2 + \vert g_s\vert^2 }{z^2}.
\label{eq:PR-PS-def}
}

\section{Power spectrum with $\delta N$ formalism}
\label{deltaN}
Using the $\delta N$-formalism~\cite{starob85,ss1,Sasaki:1998ug,lms}, we can evaluate the curvature perturbation  on the hypersurfaces of constant energy density,  $\zeta$, on super-horizon scales with the perturbation of the e-folding number $N$ defined as
\dis{
N(t_e,t_*,{\bf x}) \equiv \int_{t_*}^{t_e} H dt.
 }
 The integral is evaluated from an initial flat hyper-surface at $t=t_*$ to a final uniform density hyper-surface at $t=t_e$.
 The e-folding number $N(t_e,t_*,{\bf x})$ can be a function of the field at horizon exit at $t=t_*$ and its perturbation can be expanded in terms of the filed perturbations $\delta \phi(t_*,\bf{x})$  and $\delta \psi(t_*,\bf{x})$, 
 \dis{
 \zeta \simeq \delta N = \frac{\partial N}{\partial \phi_*} \delta \phi_* + \frac{\partial N}{\partial \psi_*} \delta \psi_*  . 
 }
 Here we assumed the slow-roll and ignored the dependence on the time derivative of the filed $\dot{\phi}$ and $\dot{\psi}$.
The field perturbation satisfies the two-point correlation function
\dis{
\VEV{\delta \phi_* ({\bf k}_1)   \delta \phi_* ({\bf k}_2)} = (2\pi)^3 \delta^{(3)}({\bf k}_1 + {\bf k}_2) \frac{2\pi^3}{k_1^3} {\mathcal P}_*(k_1), \qquad {\mathcal P}_*(k_1)\equiv \frac{H_*^2}{4\pi^2},
\label{Pstar}
}
where $H_*$ is evaluated at Hubble exit $k=a_* H_*$. The similar relation is applied to $\delta \psi_*$.
Then the power spectrum of the curvature perturbation, ${\mathcal P}_\zeta$,  is defined as
\dis{
\VEV{\zeta({\bf k}_1)  \zeta({\bf k}_2)    } =  (2\pi)^3 \delta^{(3)}({\bf k}_1 + {\bf k}_2) \frac{2\pi^3}{k_1^3} {\mathcal P}_\zeta(k_1).
\label{Pzeta}
}
From \eq{Pstar} and \eq{Pzeta}, we obtain the power spectrum of the curvature perturbation as
\dis{
{\mathcal P}_\zeta =  \frac{H_*^2}{4\pi^2} \left[  \bfrac{\partial N}{\partial \phi_*}^2 + \bfrac{\partial N}{\partial \psi_*}^2 \right].
}
In the slow-roll limit of $\phi$ field, the number of e-foldings can be written by~\cite{GarciaBellido:1995qq}
\dis{
N(\phi_*,\psi_*) = -\frac{1}{\Mp^2} \int^{\phi_e}_{\phi_*} \frac{V}{V_\phi} d\phi ,
}
where $\phi_e$ and $\psi_e$ are functions of $\phi_*$ and $\psi_*$.
The partial derivatives of the e-folding number are
\dis{
\frac{\partial N}{\partial \phi_*} =  \frac{1}{\Mp^2} \left[ \bfrac{ V}{V_\phi}_* -   \bfrac{V}{V_\phi}_e \frac{\partial\phi_e}{\partial\phi_*}\right] ,\qquad 
\frac{\partial N}{\partial \psi_*} =   - \bfrac{V}{V_\phi}_e \frac{\partial\phi_e}{\partial\psi_*}.
\label{dNdphi}
}

We note that the comoving curvature perturbation $\cR$ coincides with the perturbation on hypersurfaces of constant energy density $\zeta$ on scales far out side the horizon $k \ll aH$~\cite{Dodelson:2003ft}. In other words,
\dis{
{\mathcal P}_\zeta \approx \PR }
on super horizon scales.

\section{Reconstruction of a potential in two-field model}
\label{recon}
We are interested to reconstruct a potential that produces a power spectrum with a peak on small scales.
For this we consider a power spectrum composed of two parts, almost-scale invariant one and the other with a peak given by
\dis{
{\mathcal P}_\zeta (k) = {\mathcal P}_s (k) + {\mathcal P}_p(k),
}
with 
\dis{
 {\mathcal P}_s(k)  \equiv A_s \bfrac{k}{k_p}^{n_s-1}.
 \label{Psg}
}
Here  $k_p=0.05\mpc^{-1}$ is the pivot scale used by Planck,  and $A_s\simeq 2.0\times 10^{-9}$ and $n_s\simeq 0.96$~\cite{Akrami:2018odb}.
On large scales around CMB observation, ${\mathcal P}_s (k)$ is dominant and gives almost scale-invariant power spectrum for $k\simeq 10^{-4} - 1 \mpc^{-1}$, however on  small scales the peak spectrum is dominant  ${\mathcal P}_s \ll {\mathcal P}_p$.
In the light of the PBH formation,  we consider that ${\mathcal P}_p \gtrsim10^7{\mathcal P}_s $ around scales of the peak.

It is known that, the power spectrum that peaks at a scale $k_c=10^{12}\mpc^{-1}$ can generate stochastic background of gravitational waves which peaks in the frequency band targeted by the future interferometer LISA~\cite{Bartolo:2018evs}. By following this, in our numerical calculations, we choose to work with the parameters such that the power spectrum peaks at the scale $k_c=10^{12}\mpc^{-1}$.

In the two-field inflation models, 
we expect that the scale invariant  power spectrum $ {\mathcal P}_s$ comes from the $\phi$ field, and the one with a peak from the $\psi$ field, i.e. using $\delta N$-formalism,
\dis{
  {\mathcal P}_s = \frac{H_*^2}{4\pi^2} \bfrac{\partial N}{\partial \phi_*}^2 , \qquad 
 {\mathcal P}_p =   \frac{H_*^2}{4\pi^2}  \bfrac{\partial N}{\partial \psi_*}^2.
 \label{PsgN}
}
From these two equations, we will reconstruct the potential of two scalar fields. However,
in multi-filed case, the reconstructed potential from the power spectrum may not be unique~\cite{Andrianov:2007ua}.
In the followings, we choose a potential of the type of hybrid inflation given by 
\dis{
V=V_0 \left[1+ f(\phi) +g(\phi) h(\psi)  \right],
\label{Vtype1}
}
where $1\gg f(\phi)+g(\phi) h(\psi)$ during inflation to ensure the vacuum-domination.

The trajectories on the field space can be labelled by the integral of motion along the trajectory~\cite{GarciaBellido:1995qq}
\dis{
C = \int \frac{g(\phi)}{f_\phi(\phi)}d\phi - \int \frac{1}{h_\psi(\psi)} d\psi = F(\phi) - H(\psi),
\label{Ctraj}
}
where we defined
\dis{
 F(\phi) \equiv  \int \frac{g(\phi)}{f_\phi(\phi)} d\phi ,\qquad H(\psi) \equiv \int \frac{1}{h_\psi(\psi)} d\psi.
 \label{FH}
}
Since the integral of motion connects the field values at the horizon exit and  the end of inflation by
\dis{
F(\phi_*) - H(\psi_*)=  F(\phi_e) - H(\psi_e),
\label{FHeom}
}
we can find the relation of the partial derivatives 
\dis{
\frac{g_*}{f_{\phi_*}} =& \frac{g_e}{f_{\phi_e}} \bfrac{\partial \phi_e}{\partial \phi_*} -\frac{1}{h_{\psi_e}}  \bfrac{\partial \psi_e}{\partial \phi_*}  , \\
-\frac{1}{h_{\psi_*}} =& \frac{g_e}{f_{\psi_e}} \bfrac{\partial \phi_e}{\partial \psi_*}-\frac{1}{h_{\psi_e}}  \bfrac{\partial \psi_e}{\partial \psi_*} ,
\label{C1}
}
where we used a notation $g_*=g(\phi_*)$, $g_e=g(\phi_e)$,  $f_{\phi_*}=\left.  \frac{ d f(\phi)}{ d\phi}\right|_{\phi=\phi_*}$, and etc.
In addition to this, if we know the condition of ending inflation
\dis{
E(\phi_e,\psi_e)=0,
\label{E1}
}
then, in principle, we can obtain $\phi_e$ and $\psi_e$ in terms of $\phi_*$ and $\psi_*$ by solving \eq{FHeom} and \eq{E1} together.

From the potential in \eq{Vtype1}, the e-folding number in the slow-roll regime can be evaluated as
\dis{
N(\phi_*,\psi_*)  \simeq - \frac{1}{\Mp^2} \int_{\phi_*}^{\phi_e} \frac{d\phi}{f_{\phi}(\phi)},
\label{efold}
}
where we assumed $f_\phi(\phi)  \gg g_\phi(\phi) h(\psi)$.
Then we obtain
\dis{
\Mp^2 \frac{\partial N}{\partial \phi_*} &= -\bfrac{\partial \phi_e}{\partial \phi_*}   \frac{1}{f_{\phi_e}} + \frac{1}{f_{\phi_*}},\\\Mp^2\frac{\partial N}{\partial \psi_*} &=-\bfrac{\partial \phi_e}{\partial \psi_*} \frac{1}{f_{\phi_e}} .
\label{N1}
}
For simplicity we assume that the inflation ends by  the condition given by only $\psi_e$
\dis{
E(\phi_e,\psi_e)= E(\psi_e)=0,
}
and $\psi_e$ is independent of any $\phi_*$ and $\psi_*$.
In this case, using \eq{C1}, \eq{N1} becomes
\dis{
\Mp^2\frac{\partial N}{\partial \phi_*} &= \frac{1}{f_{\phi_*}} \left(  1-\frac{g_{*}}{g_e}\right) ,\\
\Mp^2\frac{\partial N}{\partial \psi_*} &= \frac{1}{g_e h_{\psi_*}} .
}
Now,  by matching this with the given power spectrum in \eq{PsgN}, we obtain equations
\dis{
{\mathcal P}_s  =& \frac{H_*^2}{4\pi^2 \Mp^4} \left[  \frac{1}{f_{\phi_*}} \left(  1-\frac{g_{*}}{g_e}\right)\right]^2,\\
{\mathcal P}_p = &  \frac{H_*^2}{4\pi^2 \Mp^4}   \left[  \frac{1}{g_e h_{\psi_*}}\right]^2,
\label{Pspeq}
}
where $N_*\equiv N(\phi_*,\psi_*)$. We solve these equations with the equations of motion of the fields \eq{eomN}
to reconstruct functions $f(\phi), g(\phi)$, and $h(\psi)$ in the potential.
In the following subsections, we consider two cases for them and present the power spectra which are calculated using the equations \eq{eq:ms} and \eq{eq:PR-PS-def} numerically.

\subsection{Case 1: Gaussian peak}
Here we we consider a input power spectrum with  a peak defined as:
\dis{
 {\mathcal P}_p =   
 \frac{H_*^2}{4\pi^2}  \left[ \delta + \beta e^{- \alpha (N_t-N_{tc})^2} - e^{\lambda(N_t-N_{t1})} \right]^2.
 \label{PpCase2}
 }
 Note that $N$ is defined from the end of inflation.
 However, we also use the notation of $N_t$, which is defined from some initial time of inflation with a relation $N_t=N_{\rm tot }-N$ with $N_{\rm tot}$ the e-folding number between  some initial time  and  the end.
 It is evident from the above expression that, the power {spectrum is Gaussian near $N_t=N_{tc}$ and it decreases exponentially near to the end of inflation when $N_t>N_{t1}$. 

 From this input power spectrum in \eq{PpCase2}, we try to reconstruct a potential by solving the relations in \eq{Pspeq} from $\delta N$ formalism.  By comparing both equations,  we can obtain $h_{\psi}(N_t)$ from ${\mathcal P}_p$ as
 \dis{
 h_{\psi}(N_t) = \frac{1}{\Mp^2 g_e }  \left[ \delta + \beta e^{- \alpha (N_t-N_{tc})^2} - e^{\lambda(N_t-N_{t1})} \right].
 }
In order to reconstruct a potential $h(\psi)$ which can produce the above power spectrum, we consider that the function $h_{\psi}$ in \eq{Vtype1}  is proportional to $ \psi^{n-1}$ with integer $n$ larger or equal to 2.
In this case,  $h(\psi) $ is simply
\dis{
h(\psi) = \frac{1}{n} \bfrac{\psi}{\kappa}^n,
\label{hpsi_case2}
}
where $\kappa$ is a constant with the same dimension as $\psi$. Then, \eq{hpsi_case2} directly gives the relation $\psi(N_t)$ as
\dis{ \label{eq: psi-N}
\psi (N_t) &= \psi_c  \l(\f{h_{\psi}(N_t)}{h_{\psi} (N_{tc})}\r)^{\f{1}{n-1}},
}
where $\psi_c \equiv \psi(N_{tc}) $. In the above equation, for convenience, we have rewritten $\kappa$ as $\kappa = \f{\psi_c}{(h_{\psi} (N_{tc}) \psi_c)^{1/n}}$. 
From above expression we expect that, the $\psi$ is nearly constant at initial position and slowly evolves towards a minimum and then increases. For this evolution, we find that, the second derivative of $\psi$ cannot be neglected for a short duration near the point where $\psi$ begins to roll down towards the minimum and also near the point where the first derivative of $\psi$ is zero which happens at $\psi=\psi_c$. We check this deviation of slow roll condition of $\psi$ field in our numerical calculations as well. Then, from the equation of motion, \eq{eomN}, the function $g(\phi)$ is obtained in terms of $N_t$, as
\dis{
g(\phi)
=&- \frac{H^2}{V_0 h_\psi(N_t)} \left[  \frac{d^2 \psi}{d N_t^2} + (3-\epsilon_H) \frac{d\psi}{dN_t} \right],\\
\simeq&- \frac{1}{3 h_\psi(N_t)} \left[  \frac{d^2 \psi}{d N_t^2} + 3\frac{d\psi}{dN_t} \right],
\label{gofphi}
}
where in the second line we ignored  the subdominant $\epsilon_H$, however we included the second derivative $d^2 \psi/d N_t^2$.
Using the explicit form of $\psi(N_t)$, we find
\dis{
g(\phi)=
-\f{n g_e^2 h(\phi)}{3(n-1)^2}&\bigg\{ n \l[\lambda Ex(\phi) + 2 N_d(\phi) G(\phi)\r]^2 \\
&+  (n-1) \l[\lambda (3 + \lambda) Ex(\phi) + 2 \alpha (1+N_d(\phi) (3-2 \alpha N_d(\phi))) G(\phi)\r] N_{\psi}(\phi)\bigg\}
\label{gphi_case2}
}

where
\dis{
N_{d}[\phi]\equiv & N_t(\phi)-N_{tc}\\
G(\phi) \equiv &\beta e^{-\alpha (N_t(\phi)-N_{tc})^2}\\
Ex(\phi) \equiv &e^{\lambda (N_t(\phi)-N_{t1})}
}


\begin{figure}[!t]
\begin{center}
\begin{tabular}{cc} 
 \includegraphics[width=0.38\textwidth]{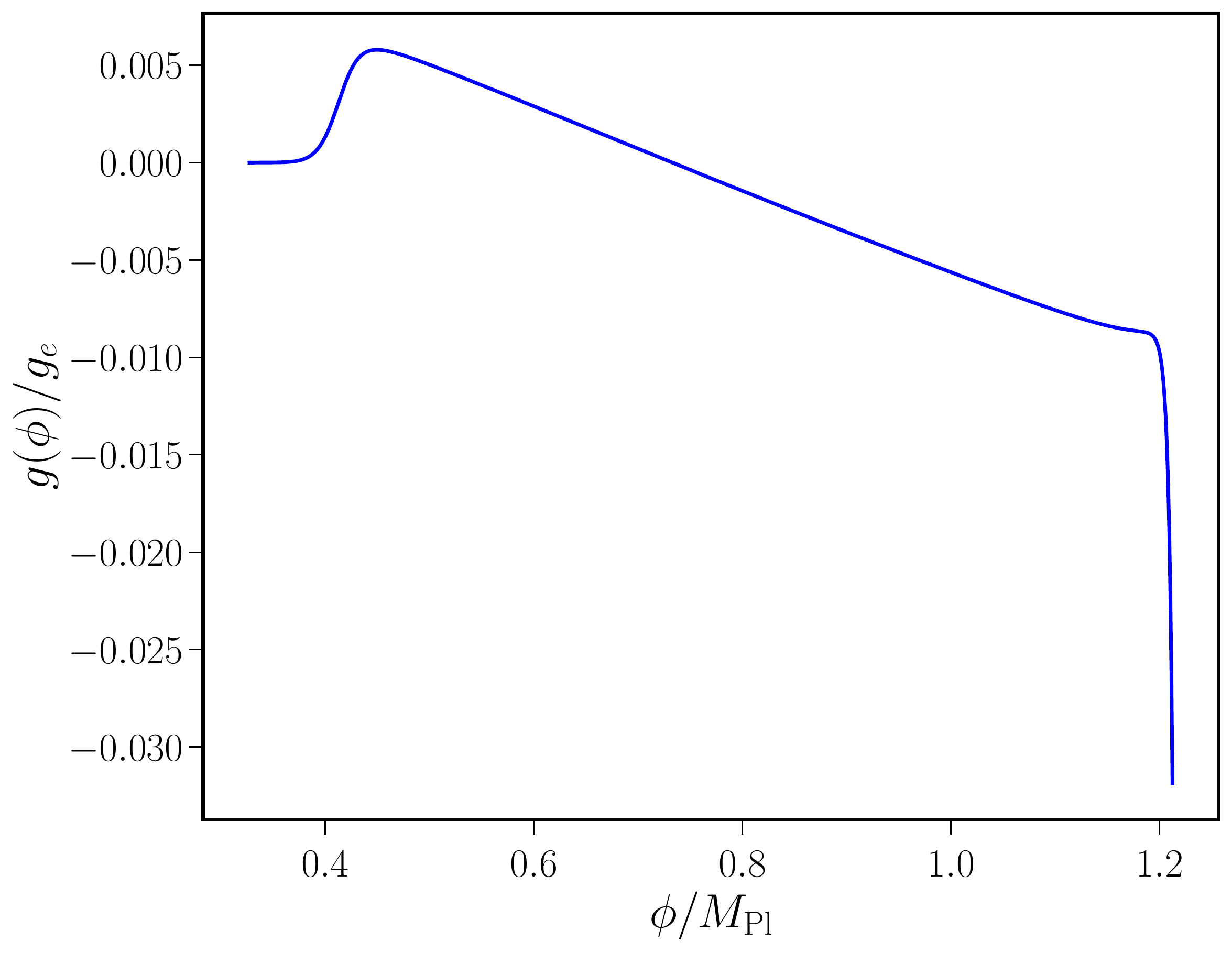}
 &
 \includegraphics[width=0.5\textwidth]{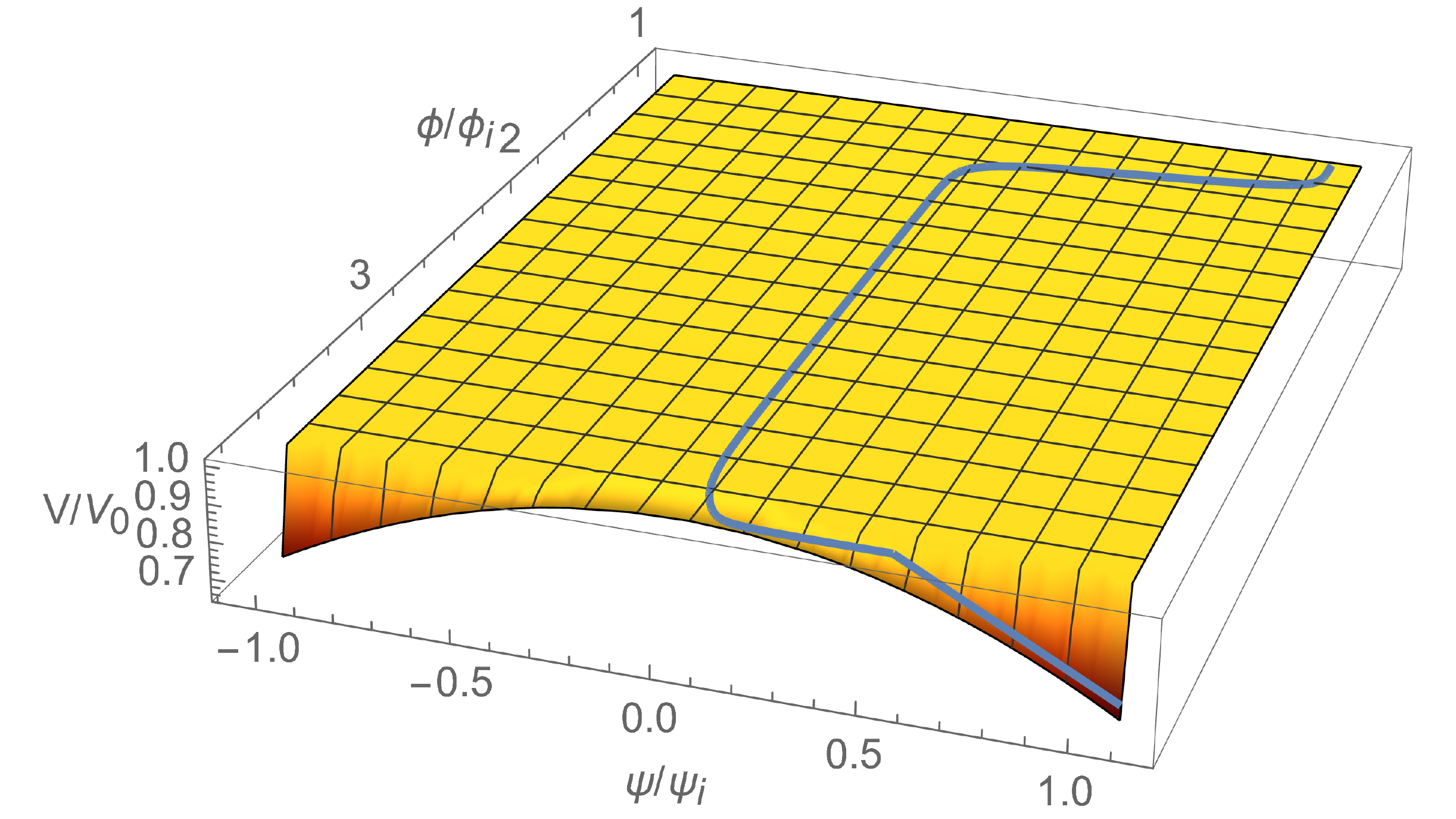}
     \end{tabular}
\end{center}
\caption{Evolution of $g(\phi) $ (left) and the trajectory of the fields on the potential $V(\phi,\psi)$ (right) for case 1 with $n=2$.}
\label{VCase2}
\end{figure}
The e-folding number $N_t$ can be replaced as a function of $\phi$ from the equation of motion in \eq{eomN} once we know the evolution of $\phi$ in terms of $N_t$. 
As a specific choice, let us use a small field inflation with a function $f(\phi)$,
\dis{
f(\phi) = - \bfrac{\phi}{\mu}^2 .
\label{fphi0}
}
By solving the slow-roll equation of motion,  we find the evolution of the field $\phi$,
\dis{
\phi = \phi_i e^{\frac{2\Mp^2}{\mu^2} (N_t-N_{t_i})}= \phi_i e^{\frac{2\Mp^2}{\mu^2} (N_i-N)},
}
and the function $g(\phi)$ is obtained by replacing $N_t$ with $\phi(N_t)$.

For the numerical calculation, we choose the power spectrum has a  peak at $k_c=10^{12}\mpc^{-1}$  corresponding to $N_{tc}=40$. We also assume that the pivot scale exit the Hubble radius at $N_{tp} = 10$. Using these information and also assuming slow roll condition, we fix,
\dis{
\mu&=10, \\
V_0 &= 24 \pi^2 2 \phi(N_p)^2 A_s/\mu^4.  
}
In addition, we choose the smallest scale relevant for the perturbations in CMB is $k_{p_2} = 1 \mpc^{-1}$ and this scale exit the Hubble radius at $N_{tp_2}= N_{tp} + \log{\f{k_{p_2}}{k_p}}$. As we have mentioned earlier, on  small scales the peak spectrum is dominant, \ie ${\mathcal P}_s \ll {\mathcal P}_p$ and also ${\mathcal P}_p \gtrsim10^7{\mathcal P}_s $ around scales of the peak. These information leads to the constraints
\dis{
\delta<&\f{\mu^2}{2\phi(N_{tp_2})}, \ \\
\beta>& \f{\mu^2\sqrt{10^7}}{2  \phi(N_{tp})}. \ \\
}
For the numerical calculations, we choose to work with 
\dis{
\alpha= 1/75, \quad
\delta= 12, \quad
\beta = 4 \times 10^{5}.
} The remaining parameters are the initial field values used at $N_{t}=0$ which are fixed as
\dis{
\phi_i=\phi(N_{tp}) e^{-2 N_{tp}/\mu^2}, \nn  \qquad  \psi_i =10^{-3}.
}

\begin{figure}[!t]
\begin{center}
\begin{tabular}{c} 
 \includegraphics[width=0.5\textwidth]{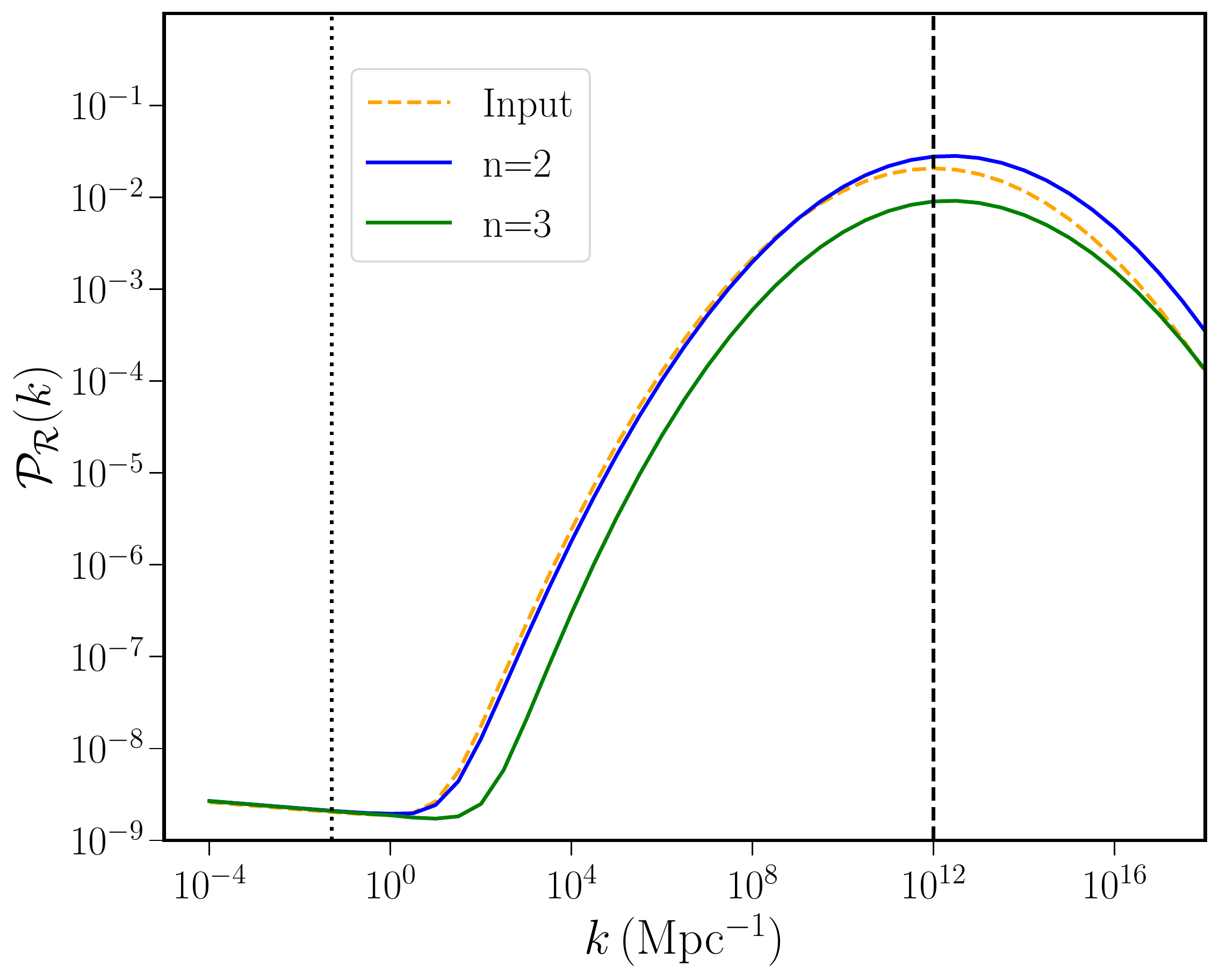}
      \end{tabular}
\end{center}
\caption{Left: The power spectrums calculated numerically from the potential in Case 1 with  $n=2$ (blue), $n=3$ (green), and the input power spectrum (orange). We used $\alpha=1/75, \beta=4\times 10^5, \delta =12 $, $N_{tc}=40$ and $N_{t1} = 65$, with $\lambda =5$.}
\label{PCase2}
\end{figure}

In Fig.~\ref{VCase2} (Left),  we show the function $g(\phi)$ for the function $f(\phi)$ used in \eq{fphi0}. The explicit form of the potential is obtained substituting \eq{hpsi_case2} and \eq{gphi_case2} in \eq{Vtype1}, which is complicated and not shown here.
Instead, in Fig.~\ref{VCase2} (Right), we show the potential and the trajectory of the fields in the plane of  ($\phi/\phi_i, \psi/\psi_i$).
It is interesting to note that, during the initial stages of inflation, the function $g(\phi)$ is very small. This is expected since the potential is dominated by the field $\phi$ alone. In the case of $n=2$ the square of the mass of $\psi$ is proportional to the function $g(\phi)$. Around $N_t =N_{tc}$, $g(\phi)$ is linear in $N_t$ and changes its sign from positive to negative. This means that the field $\psi$ becomes tachyonic. Finally, the square of mass decreases exponentially and this leads inflation to end when the $\epsilon_H=1$.

In Fig.~\ref{PCase2}, we show the power spectrums for this case: the input power spectrum (orange dashed) and the power spectrums from the reconstructed potential with  $n=2$ (blue solid), $n=3$ (green solid).
We can see that the input power spectrum matches the power spectrum obtained numerically quite well within an error of $10\%$.
\begin{figure}[!t]
\begin{center}
\begin{tabular}{cc} 
 \includegraphics[width=0.38\textwidth]{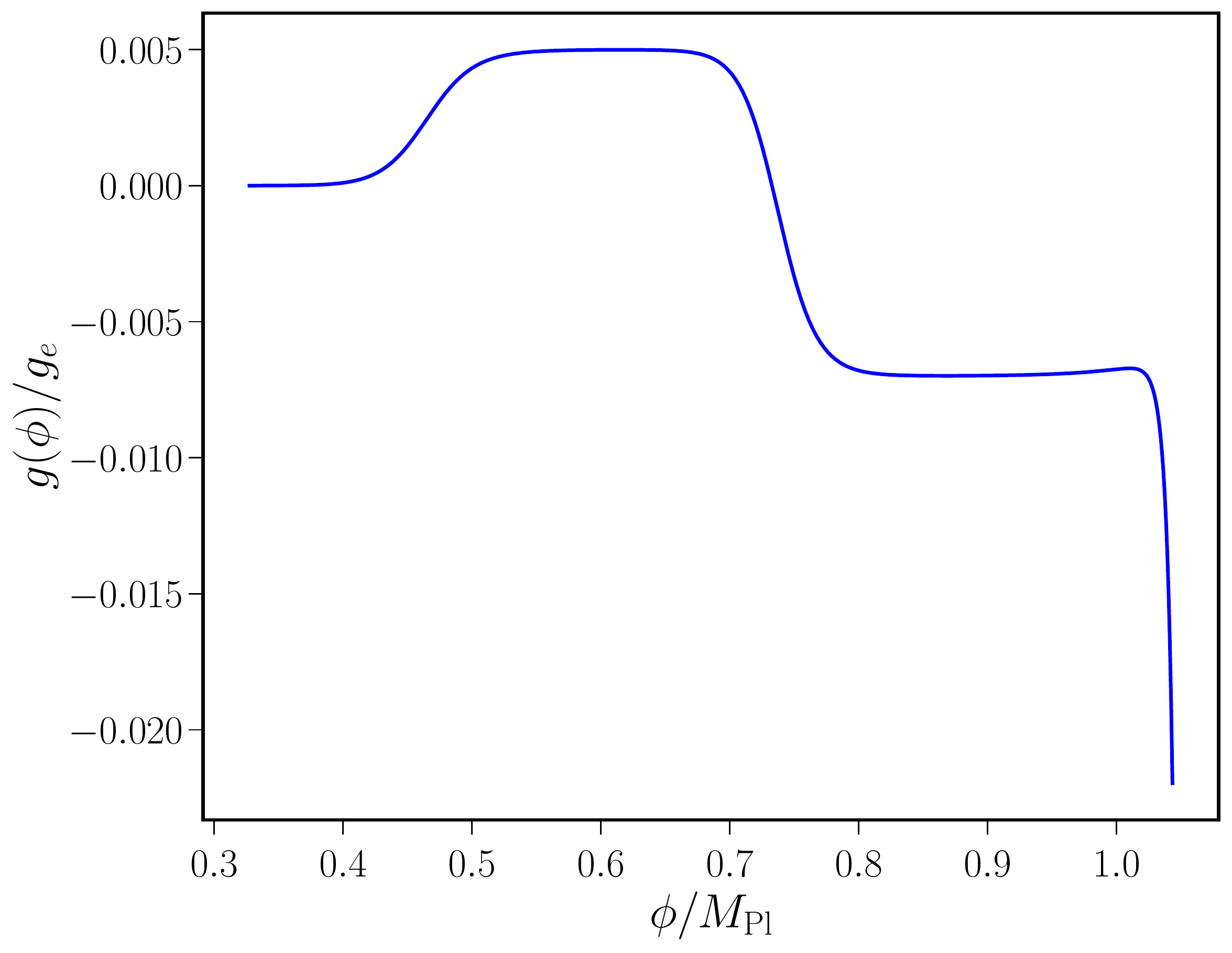}
 &
 \includegraphics[width=0.5\textwidth]{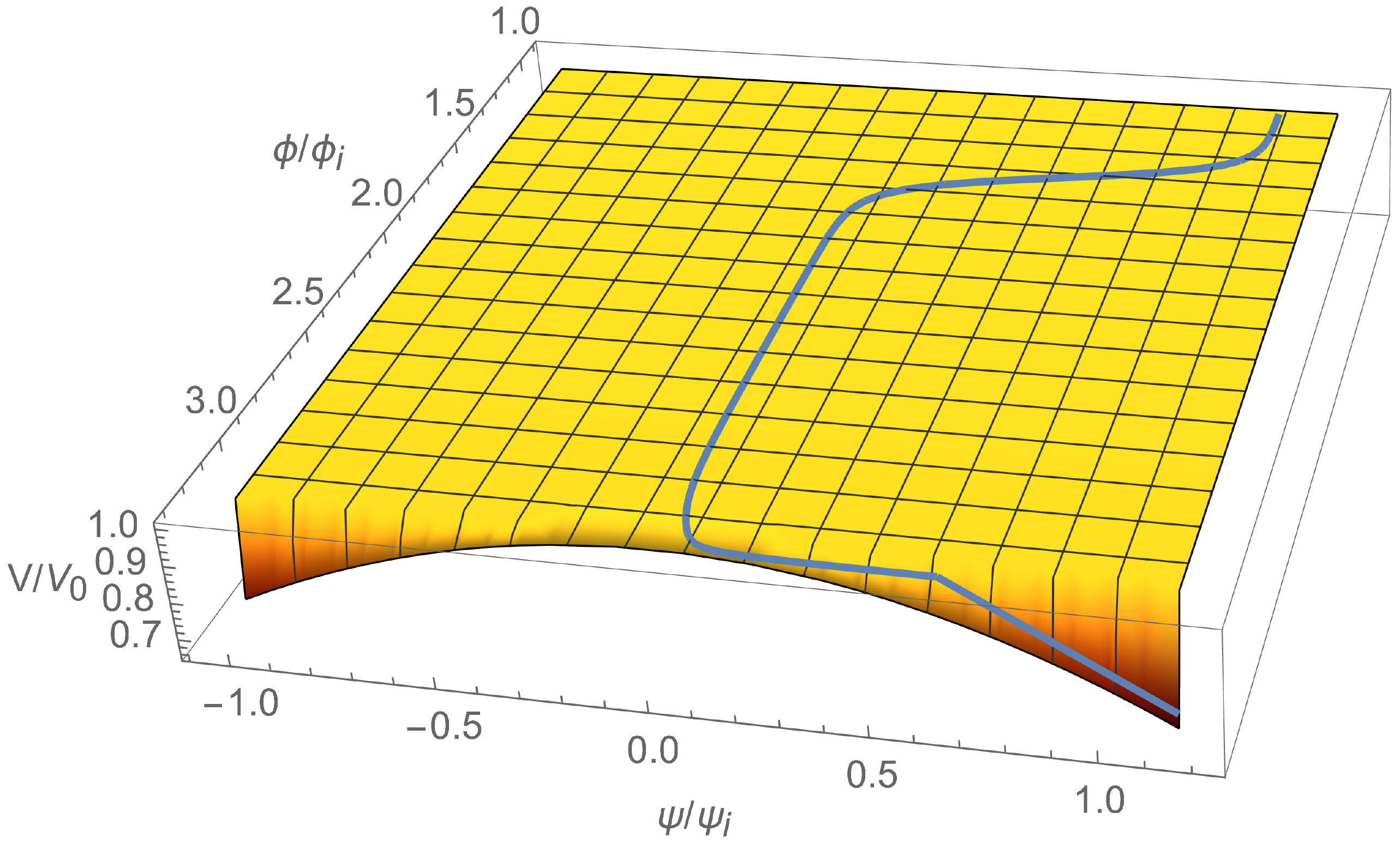}
     \end{tabular}
\end{center}
\caption{Evolution of $g(\phi) $ (left) and the trajectory of the fields on the potential $V(\phi,\psi)$ (right) for case 2 with $n=2$ . We used the same parameters as in Fig.~\ref{PCase2} except 
$\gamma= 1/2, N_{t1} = 57$, and $\lambda =3$.}
\label{gophi-sech}
\end{figure}
\begin{figure}[!t]
\begin{center}
 \includegraphics[width=0.5\textwidth]{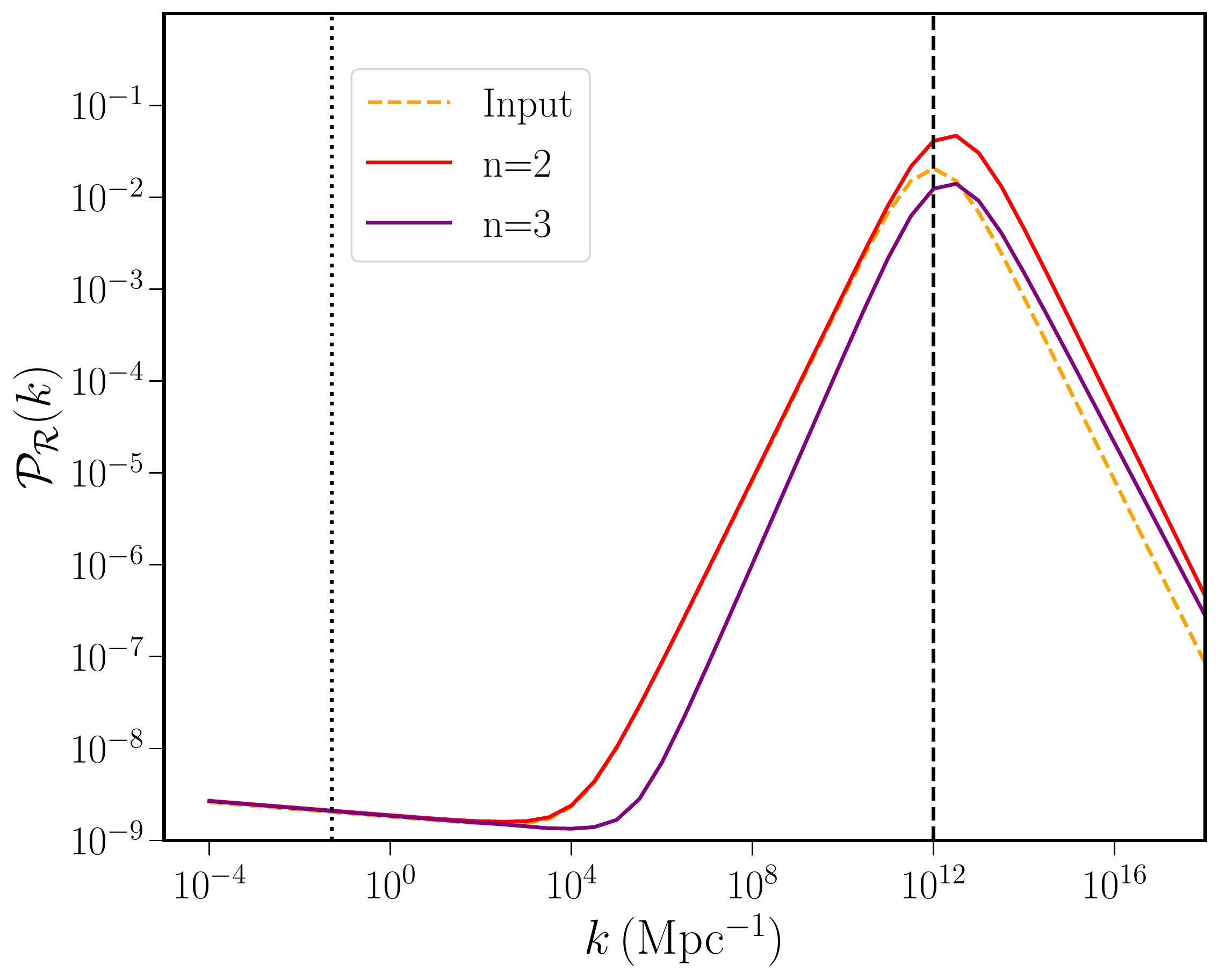}
\end{center}
\caption{The power spectrums calculated numerically from the reconstructed potential in Case 2 with $n=2$ (red), $n=3$ (purple) and the input power spectrum (orange). We used the same parameters as in Fig.~\ref{PCase2} except $\gamma= 1/2, N_{t1} = 57$, and $\lambda =3$.}
\label{PCase3}
\end{figure}
\subsection{Case 2 : Hyperbolic peak}
In the previous case, we have been able to reconstruct the potential by considering the Gaussian power spectrum and the function $h_{\psi} \propto \psi^{n}$. Let us now try to reconstruct the potential by assuming the same $h_{\psi}$ but a different type of peak power spectrum which is given by
\dis{
 {\mathcal P}_p =  
 \frac{H_*^2}{4\pi^2}  \left[\delta + \beta {\rm sech}\l[\gamma (N_t-N_{tc})\r] - e^{\lambda(N_t-N_{t1})}\right]^2.\\
 \label{PpCase3}
 }
The task is to reconstruct $g(\phi)$ by using the  \eq{gofphi}. Following the same method used in case 1, it is straight forward to obtain $g(\phi)$ as
\dis{
g(\phi) = 
 -\f{n g_e^2 h(\phi)}{3(n-1)^2} & \bigg\{n \lambda^2 Ex(\phi)^2 + \lambda Ex(\phi)\l[(n-1) (3+\lambda) N_{\psi}(\phi) + 2 n \beta \gamma S[\phi] T[\phi] \r]  \\
 &+ \gamma \beta S(\phi) \l[ n \beta \gamma S(\phi) T(\phi)^2 + (n-1) N_{\psi}(\phi) (2 \gamma S(\phi)^2 - \gamma + 3T(\phi))\r]\bigg\}
}
where 
\dis{ S(\phi) &= \f{2}{(\phi/\phi_c)^{\gamma \mu^2/2} + (\phi/\phi_c)^{-\gamma \mu^2/2}}, \\
        T(\phi) & = \f{(\phi/\phi_c)^{\gamma \mu^2} - (\phi/\phi_c)^{-\gamma \mu^2}}{(\phi/\phi_c)^{\gamma \mu^2} + (\phi/\phi_c)^{-\gamma \mu^2}}.
}
In Fig. \ref{gophi-sech}, we show the evolution of $g(\phi) $ and the trajectory of the fields on the potential $V(\phi,\psi)$ for case 2 with $n=2$. As one can see from this figure, the overall behavior of the function $g(\phi)$ and the evolution of the inflationary trajectory are similar as in the case 1.
In Fig.\ref{PCase3}, we show the power spectrum for the reconstructed potential in Case 2 with $n=2$ (red), $n=3$ (purple)with the input power spectrum (orange). Here we used  $\gamma= 1/2, N_{t1} = 57$, and $\lambda =3$. It is clear from the figure that the given spectrum match the numerical results very well.

\section{Discussion}
\label{con}
The primordial black hole can be produced from the enhanced primordial power spectrum of the curvature perturbation on small scales, and may play as dark matter. In the literature, there have been several trials to calculate an enhanced power spectrum from a given potential in single field or multi-field inflation models. In this paper,  however, we suggest new  method to reconstruct a potential from a given power spectrum. With this way, we could find potentials that have a peak on small scales, which is large enough to generate primordial black holes.

In this work, we used the input power spectrum composed of a nearly scale-invariant one on large scales and the other with a peak on small scales. To reconstruct a potential with two canonical scalar fields with $\phi$ and $\psi$, we have used a hybrid type potential. 
Guided by the $\delta N$ formalism,  we matched each power spectrum to the contribution from each field and solved them.
We have been able to solve them numerically and could reconstruct a potential. We evaluated the scalar power spectra in reconstructed models and confirmed that  the resulting spectra are quite compatible with the input spectrum.
 
In the reconstructed models, the scale-invariant power spectrum is generated when the field $\phi$ is dominant while the field $\psi$ is nearly constant. After this period, the field $\psi$ starts moving towards its minimum and  then bounces at a critical value to increase. 
During this bounce, the power spectrum is dominated by the $\psi$ field and  the enhance peak is generated. 

We should also mention that, though the correct form of the power spectrum for small scales is yet to be understood, for illustration, we have chosen two types of power spectrum, Gaussian and hyperbolic. It is also important to note that, in this work, we have focussed on constructing small-field hybrid type of potentials. We believe that, using our methods discussed in this work, one can explore more complex models beyond the models we have constructed here.

\begin{acknowledgments}
K.-Y.C. and R.N.Raveendran were supported by the National Research Foundation of Korea (NRF) grant funded by the Korea government (MEST) (NRF-2019R1A2B5B01070181). S. Kang was supported by Korea Initiative for fostering University of Research and Innovation Program of the National Research Foundation (NRF) funded by the Korean government (MSIT) (No.2020M3H1A1077095).
\end{acknowledgments}



\begin{thebibliography}{99}
  
\bibitem{Akrami:2018odb}
Y.~Akrami \textit{et al.} [Planck],
Astron. Astrophys. \textbf{641} (2020), A10
doi:10.1051/0004-6361/201833887
[arXiv:1807.06211 [astro-ph.CO]].


\bibitem{Bringmann:2011ut}
T.~Bringmann, P.~Scott and Y.~Akrami,
Phys. Rev. D \textbf{85} (2012), 125027
doi:10.1103/PhysRevD.85.125027
[arXiv:1110.2484 [astro-ph.CO]].

\bibitem{Choi:2015yma}
K.~Y.~Choi, J.~O.~Gong and C.~S.~Shin,
Phys. Rev. Lett. \textbf{115} (2015) no.21, 211302
doi:10.1103/PhysRevLett.115.211302
[arXiv:1507.03871 [astro-ph.CO]].



\bibitem{Zeldovich:1967lct}
Y.~B.~;.~N.~Zel'dovich, I.~D.,
Soviet Astron. AJ (Engl. Transl. ), \textbf{10} (1967), 602


\bibitem{Hawking:1971ei}
S.~Hawking,
Mon. Not. Roy. Astron. Soc. \textbf{152} (1971), 75

\bibitem{Carr:1975qj}
B.~J.~Carr,
Astrophys. J. \textbf{201} (1975), 1-19
doi:10.1086/153853

\bibitem{Page:1976df}
D.~N.~Page,
Phys. Rev. D \textbf{13} (1976), 198-206
doi:10.1103/PhysRevD.13.198


\bibitem{Green:2020jor}
A.~M.~Green and B.~J.~Kavanagh,
[arXiv:2007.10722 [astro-ph.CO]].


\bibitem{Ivanov:1994pa}
P.~Ivanov, P.~Naselsky and I.~Novikov,
Phys. Rev. D \textbf{50} (1994), 7173-7178
doi:10.1103/PhysRevD.50.7173

\bibitem{Stewart:1996ey}
E.~D.~Stewart,
Phys. Lett. B \textbf{391} (1997), 34-38
doi:10.1016/S0370-2693(96)01458-X
[arXiv:hep-ph/9606241 [hep-ph]].

\bibitem{Kohri:2007gq}
K.~Kohri, C.~M.~Lin and D.~H.~Lyth,
JCAP \textbf{12} (2007), 004
doi:10.1088/1475-7516/2007/12/004
[arXiv:0707.3826 [hep-ph]].


\bibitem{Kohri:2007qn}
K.~Kohri, D.~H.~Lyth and A.~Melchiorri,
JCAP \textbf{04} (2008), 038
doi:10.1088/1475-7516/2008/04/038
[arXiv:0711.5006 [hep-ph]].


\bibitem{Alabidi:2009bk}
L.~Alabidi and K.~Kohri,
Phys. Rev. D \textbf{80} (2009), 063511
doi:10.1103/PhysRevD.80.063511
[arXiv:0906.1398 [astro-ph.CO]].


\bibitem{Garcia-Bellido:2017mdw}
J.~Garcia-Bellido and E.~Ruiz Morales,
Phys. Dark Univ. \textbf{18} (2017), 47-54
doi:10.1016/j.dark.2017.09.007
[arXiv:1702.03901 [astro-ph.CO]].

\bibitem{Germani:2017bcs}
C.~Germani and T.~Prokopec,
Phys. Dark Univ. \textbf{18} (2017), 6-10
doi:10.1016/j.dark.2017.09.001
[arXiv:1706.04226 [astro-ph.CO]].

\bibitem{Motohashi:2017kbs}
H.~Motohashi and W.~Hu,
Phys. Rev. D \textbf{96} (2017) no.6, 063503
doi:10.1103/PhysRevD.96.063503
[arXiv:1706.06784 [astro-ph.CO]].

\bibitem{Ballesteros:2017fsr}
G.~Ballesteros and M.~Taoso,
Phys. Rev. D \textbf{97} (2018) no.2, 023501
doi:10.1103/PhysRevD.97.023501
[arXiv:1709.05565 [hep-ph]].

\bibitem{Hertzberg:2017dkh}
M.~P.~Hertzberg and M.~Yamada,
Phys. Rev. D \textbf{97} (2018) no.8, 083509
doi:10.1103/PhysRevD.97.083509
[arXiv:1712.09750 [astro-ph.CO]].

\bibitem{Fumagalli:2020adf}
J.~Fumagalli, S.~Renaux-Petel, J.~W.~Ronayne and L.~T.~Witkowski,
[arXiv:2004.08369 [hep-th]].

\bibitem{Yokoyama:1995ex}
J.~Yokoyama,
Astron. Astrophys. \textbf{318} (1997), 673
[arXiv:astro-ph/9509027 [astro-ph]].

\bibitem{Kawasaki:1997ju}
M.~Kawasaki, N.~Sugiyama and T.~Yanagida,
Phys. Rev. D \textbf{57} (1998), 6050-6056
doi:10.1103/PhysRevD.57.6050
[arXiv:hep-ph/9710259 [hep-ph]].

\bibitem{Yokoyama:1998pt}
J.~Yokoyama,
Phys. Rev. D \textbf{58} (1998), 083510
doi:10.1103/PhysRevD.58.083510
[arXiv:astro-ph/9802357 [astro-ph]].


\bibitem{Kawasaki:2012wr}
M.~Kawasaki, N.~Kitajima and T.~T.~Yanagida,
Phys. Rev. D \textbf{87} (2013) no.6, 063519
doi:10.1103/PhysRevD.87.063519
[arXiv:1207.2550 [hep-ph]].

\bibitem{Clesse:2015wea}
S.~Clesse and J.~Garc\'\i{}a-Bellido,
Phys. Rev. D \textbf{92} (2015) no.2, 023524
doi:10.1103/PhysRevD.92.023524
[arXiv:1501.07565 [astro-ph.CO]].


\bibitem{Braglia:2020eai}
M.~Braglia, D.~K.~Hazra, F.~Finelli, G.~F.~Smoot, L.~Sriramkumar and A.~A.~Starobinsky,
JCAP \textbf{08} (2020), 001
doi:10.1088/1475-7516/2020/08/001
[arXiv:2005.02895 [astro-ph.CO]].

\bibitem{Palma:2020ejf}
G.~A.~Palma, S.~Sypsas and C.~Zenteno,
Phys. Rev. Lett. \textbf{125} (2020) no.12, 121301
doi:10.1103/PhysRevLett.125.121301
[arXiv:2004.06106 [astro-ph.CO]].


\bibitem{Gordon:2000hv}
C.~Gordon, D.~Wands, B.~A.~Bassett and R.~Maartens,
Phys. Rev. D \textbf{63} (2000), 023506
doi:10.1103/PhysRevD.63.023506
[arXiv:astro-ph/0009131 [astro-ph]].


\bibitem{Lalak:2007vi}
Z.~Lalak, D.~Langlois, S.~Pokorski and K.~Turzynski,
JCAP \textbf{07} (2007), 014
doi:10.1088/1475-7516/2007/07/014
[arXiv:0704.0212 [hep-th]].



\bibitem{GarciaBellido:1995qq}
J.~Garcia-Bellido and D.~Wands,
Phys. Rev. D \textbf{53} (1996), 5437-5445
doi:10.1103/PhysRevD.53.5437
[arXiv:astro-ph/9511029 [astro-ph]].


\bibitem{starob85}
 A.~A.~Starobinsky,
  JETP Lett.\  {\bf 42}, 152 (1985)
  [Pisma Zh.\ Eksp.\ Teor.\ Fiz.\  {\bf 42}, 124 (1985)].

\bibitem{ss1}
 M.~Sasaki   and E.~D.~Stewart,
  Prog.\ Theor.\ Phys.\  {\bf 95} (1996) 71
[arXiv:astro-ph/9507001].

\bibitem{Sasaki:1998ug}
 M.~Sasaki and T.~Tanaka,
 Prog.\ Theor.\ Phys.\  {\bf 99}, 763 (1998)
 [arXiv:gr-qc/9801017].

\bibitem{lms}
D.~H.~Lyth, K.~A.~Malik and M.~Sasaki,
JCAP {\bf 0505}, 004 (2005)
[arXiv:astro-ph/0411220].


\bibitem{Dodelson:2003ft}
  S.~Dodelson,
Amsterdam, Netherlands: Academic Pr. (2003) 440 p

\bibitem{Bartolo:2018evs}
N.~Bartolo, V.~De Luca, G.~Franciolini, A.~Lewis, M.~Peloso and A.~Riotto,
Phys. Rev. Lett. \textbf{122} (2019) no.21, 211301
doi:10.1103/PhysRevLett.122.211301
[arXiv:1810.12218 [astro-ph.CO]].



\bibitem{Andrianov:2007ua}
A.~A.~Andrianov, F.~Cannata, A.~Y.~Kamenshchik and D.~Regoli,
JCAP \textbf{02} (2008), 015
doi:10.1088/1475-7516/2008/02/015
[arXiv:0711.4300 [gr-qc]].





%
\bibitem{Lyth:2005fi}
D.~H.~Lyth and Y.~Rodriguez,
Phys. Rev. Lett. \textbf{95} (2005), 121302
doi:10.1103/PhysRevLett.95.121302
[arXiv:astro-ph/0504045 [astro-ph]].
%
\bibitem{Lyth:1996kt}
D.~H.~Lyth and E.~D.~Stewart,
Phys. Rev. D \textbf{54} (1996), 7186-7190
doi:10.1103/PhysRevD.54.7186
[arXiv:hep-ph/9606412 [hep-ph]].
%




  \end{thebibliography}
\end{document}